\documentclass{aa}

\usepackage{graphicx}

\usepackage{txfonts} 

\begin{document}

\title{The gravitational stratification of multifluid and multispecies plasma} 

     \author{F. Zhang\inst{1,2}    
          \and
          J. Mart\'inez-Sykora\inst{3,4,1,2}  
          \and
          Q. M. Wargnier\inst{5}
          \and 
          V. H. Hansteen\inst{1,2,3} 
}

   \institute{Institute of Theoretical Astrophysics, University of Oslo, PO Box 1029 Blindern, 0315 Oslo, Norway  \\
              \email{fan.zhang@astro.uio.no} 
\and 
Rosseland Centre for Solar Physics, University of Oslo, PO Box 1029 Blindern, 0315 Oslo, Norway 
\and
 SETI Institute, 339 Bernardo Ave, Mountain View, CA 94035, USA
\and
Lockheed Martin Solar and Astrophysics Laboratory, 3251 Hanover St, Palo Alto, CA 94304, USA
\and
Green 14 AB, Brinellvägen 25, Stockholm, Sweden
             }

  \abstract 
   {The solar atmosphere is gravitationally stratified and consists of several layers at temperatures that vary by several orders of magnitude. Consequently, the solar atmospheric plasma changes from weakly ionized in the photosphere, partially ionized in the chromosphere, and to fully ionized in the corona. However, {integrating ionization and recombination processes into multifluid solar plasma models with gravitational stratification continues to be a nontrivial task.}} 
   {We intend to provide a method for constructing multifluid+multispecies (MFMS) gravitational stratification that satisfies the ionization equilibrium and hydrostatic equilibrium at the same time, avoiding causing nonphysical disturbances and numerical instability due to the initial {imbalances}.}
{We assume that collisional interactions between fluids are sufficient for coupling all fluids when there is no high-frequency external driving force imposed. Ionization fractions can be (I) calculated assuming ionization in statistical equilibrium at any given temperature or (II) extracted from other atmospheric models. A simple numerical integration routine would then be used to construct MFMS gravitational stratifications.}
   {{The gravitational stratification in hydrostatic equilibrium can be constructed using the present numerical integration routine} with any given ionization fractions of multispecies plasmas. Meanwhile, without any dynamic driving force, fluid decoupling is initiated, particularly in the transition region of the constructed stratification, while the total velocity of all fluids remains at the level of zero.}
   {A gravitational stratification constructed using the present routine can be used in MFMS models to study specific dynamics without being affected by the initial {imbalances}.}
   {}
 \titlerunning{The gravitational stratification of multifluid and multispecies plasma}
   \maketitle

\section{Introduction}

Solar and stellar atmospheres are highly complex and dynamic, and to numerically model and analyze them, we generally need to make various assumptions and simplifications. An important and frequently applied approach is to assume a 1D "background" atmosphere in a static gravitational stratification, following 
\begin{equation} \label{eq:Hydro}
\frac{\text{d} P}{\text{d} z}=-\varrho g,
\end{equation}
\noindent where the total pressure includes thermal pressure of gas plasma and magnetic pressure, namely, $P=p^{\text{g(as)}}+p^{\text{m(agnetic)}}$, $\varrho$ is the plasma density, $g$ is the gravitational acceleration, and $z$ is the height. This balance between the total pressure gradient and gravity is assumed to hold in the atmospheric model along the $z$ direction. We note that in the solar atmosphere, the mean-free path of particles is much smaller than the characteristic length scales of interest. Therefore, the plasma can be assumed to be in Maxwellian equilibrium and a fluid description e.g., magnetohydrodynamics {{(MHD)}}, is appropriate. {Consequently}, the stationary momentum equation reduces to the hydrostatic balance as written in Eq.~(\ref{eq:Hydro}). 
Models using this assumption have been extensively used to investigate, for example, MHD waves in the solar atmosphere  \citep{Khomenko_2015,VanDoorsselaere2020}. 
 
However, this common practice encounters challenges when considering multifluid (two or more fluids) MHD modeling of, for instance, the low solar atmosphere, where the thermodynamic conditions do not permit a single-fluid description, and instead require a set of coupled governing equations for the ionized and neutral components, commonly referred to as multifluid MHD equations \citep{Khomenko2016,Ballester2018}.       {In  multifluid models}, the balance between pressure gradient and gravity is typically assumed to be independent for each fluid. More specifically, when constructing an initially static atmospheric model, it is frequently assumed that each of the $n$ fluids is in an independent hydrostatic equilibrium, 
\begin{equation}  \label{eq:Hydro_n}
\frac{\text{d} P_i}{\text{d} z}=-\varrho_i g, \quad i=1, ..., n,
\end{equation}
\noindent where the thermal pressure, $p_i^{\text{g}}$, of the $i$-th fluid includes the thermal pressure of the free electrons produced when the $i$-th fluid underwent ionization. For example, in various pure-hydrogen ion-neutral two-fluid models \citep{Laguna2016,PopescuBraileanu2019,Wojcik2020,PopescuBraileanu_2023}, the equations of state of neutral and ionized hydrogen are, respectively,
\begin{equation}   \label{eq:EoS_1}
p^{\text{g}}_{\text{n(eutral)}}=N_{\text{n}}k_{\text{B}}T_{\text{n}},
\end{equation}
\noindent and
\begin{equation}  \label{eq:EoS_2}
p^{\text{g}}_{\text{i(on)}}=N_{\text{i}}k_{\text{B}}T_{\text{i}}+N_{\text{e(lectron)}}k_{\text{B}}T_{\text{e}}=2N_{\text{i}}k_{\text{B}}T_{\text{i}}, 
\end{equation}
\noindent where $N$ and $T$ denote the number densities and temperatures, with subscripts i, n, and e indicating ionized hydrogen, neutral hydrogen, and free electrons, respectively; $k_{\text{B}}$ is the Boltzmann constant.       {These two independent equations of state lead to different {and independent} scale heights of neutral and ionized hydrogen.} Moreover, in these models, the description of free electrons is simplified, ignoring the thermal force, which is feasible for modeling the low solar atmosphere, which can be assumed to be nearly isothermal \citep{Hansteen_1997}. In addition, Eq.~(\ref{eq:EoS_2}) assumes that the ion temperature and electron temperature are the same. This assumption simplifies the numerical model but is not a necessary constraint for the solar atmosphere.

Directly using Eq.~(\ref{eq:Hydro_n}) to calculate a multifluid gravitational stratification causes practical issues. 
As the scale heights are {independent} for different fluids, the resulting abundances at certain heights may differ by orders of magnitude from atmospheric models that holistically combine the contributions of necessary physical mechanisms \citep{Vernazza1981,Fontenla_1993,Avrett2008}. Such differences then change plasma properties and thus the results from numerical models \citep{Zhang2021}. Moreover, what Eq.~(\ref{eq:Hydro}) or (\ref{eq:Hydro_n}) describes is a boundary value problem that depends on the values of $\varrho_i(z_0)$,  where $z_0$ is the reference height and temperature only plays a role through the equations of state. Thus, the resulting stratified plasma is largely not in ionization equilibrium, which is directly related to the local temperature, and static atmospheric models given by Eq.~(\ref{eq:Hydro_n}) cannot be used to start simulations with ionization and recombination, which will drastically change the energy balance. Using Eq.~(\ref{eq:Hydro_n}) also implies the assumption that collisional interactions between different fluids do not affect the static equilibrium. Intuitively, however, one may still expect collisional interactions between fluids, although such interactions might not always be sufficient to fully couple all fluids \citep{Hansteen_1997}. {Physically, collisional processes are also important in ionization equilibrium \citep{House_1964}.} Therefore, it is necessary to revise the assumption of independent hydrostatic equilibria for different fluids.

Obtaining an initial atmospheric model that is not only in hydrostatic equilibrium, but also in ionization equilibrium is not a trivial endeavor. Thus far, only a small number of two-fluid low solar atmospheric models \citep{Maneva_2017,Zhang2021,Niedziela_2024,Kraskiewicz_2025} have included both equilibria. Indeed, it is possible to run a long(er) physical-time simulation to reach a steady state \citep{Brchnelova2023a}, where the initial imbalances eventually disappear. However, the numerical solution would depend on the chosen initial conditions and it is difficult to constrain the solution to obtain a specific atmospheric composition.       {In addition, this strategy} can also be computationally intensive. Therefore, to obtain an initial static stratification for a multifluid+multispecies (MFMS)  model that could include various heavier elements \citep{Wargnier_2023,Sykora_2023} in addition to hydrogen, an alternative solution is needed.
 
To construct an MFMS hydrostatic equilibrium atmospheric model, we assume a static (or steady) configuration in the absence of external disturbances or force imbalances, such that the characteristic timescale of the equilibrium state is infinitely larger than the collisional timescales. Therefore, we may assume that all the $n$ fluids are coupled by collisional interactions (or other mixing processes),\footnote{This assumption is, however, not valid higher up in the solar atmosphere and the solar wind, where the plasma gradually becomes collisionless and different fluids and species become decoupled \citep{Hansteen_1994,Hansteen_1995}.}approximately acting as a single fluid \citep{Khomenko2012,Martinez-Sykora_2012} in the static state. Then, the gravitational stratification can be  calculated in a straightforward way. \citet{Zhang2021} used this idea to provide initial equilibria for pure-hydrogen two-fluid plasmas under an isothermal condition. The idea can be extended to MFMS modeling by using a simple numerical integration routine, which is the focus of this work. Below, we begin by explaining this simple idea using an isothermal example similar to those of \citet{Zhang2021} and afterwards more realistic temperature stratification and ionization are taken into account. We also address the essential physical and numerical consequences of using our model. Additional details are introduced in the appendix.

\section{The basics} \label{sec:example}

\subsection{A coupled approach}

Considering multifluid plasmas, we calculated two different types of stratification, as discussed below: 
\begin{itemize}
    \item[(I)] {Pure hydrostatic equilibrium (pHE):} Each fluid is treated independently and satisfies its own hydrostatic equilibrium condition, as given by Eq.~(\ref{eq:Hydro_n}).
    \item[(II)] {Coupled hydrostatic equilibrium (cHE):} The hydrostatic equilibrium is established through coupling between different fluids, such that their equilibrium states are mutually dependent.
\end{itemize}

We note that because constructing the pHE is straightforward, we only offers details on the cHE below. As we assume that in the cHE stratification all fluids are coupled by collisional interactions, at each height, all fluids can be described by their bulk motion. Thus, we define a mean atomic mass, which is expressed as
\begin{equation} \label{eq:mt}
\overline{m}_{\text{T}}=\frac{\sum_{i=1}^{n+1} N_im_i}{\sum_{i=1}^{n+1} N_i},
\end{equation}
\noindent where $m_i$ and $N_i$ are, respectively the atomic mass and the number density of the $i$-th fluid.       In particular, we should emphasize that free electrons should be considered (i.e., the $n$+1-th fluid) when calculating the mean atomic mass,  {without breaking charge neutrality}, while the electron mass is assumed to be zero.  While assuming the ideal gas law, we have the equivalent equation of state, 
\begin{equation} \label{eq:pT}
p_{\text{T}}^{\text{g}}=N_{\text{T}}k_{\text{B}}T,
\end{equation}
\noindent where $p_{\text{T}}^{\text{g}}$ is the  plasma total thermal pressure and $N_{\text{T}}=\sum_{i=1}^{n+1} N_i$ is the total number density.       {We can also write out the partial pressures as}
\begin{equation} \label{eq:partialND}
p_{i}^{\text{g}}=N_{i}k_{\text{B}}T, \quad i=1, ..., n+1,
\end{equation}
\noindent   by assuming that the temperature is the same for all fluids.
Finally, Eq.~(\ref{eq:Hydro}) can be rewritten for the MFMS model, giving
\begin{equation} \label{eq:P_MFMS}
\frac{\text{d} p_{\text{T}}^{\text{g}}}{\text{d} z}=-N_{\text{T}}\overline{m}_{\text{T}}g=-\frac{p_{\text{T}}^{\text{g}}\overline{m}_{\text{T}}g}{k_{\text{B}}T},
\end{equation}

\noindent  {without accounting for the contribution of magnetic pressure. Magnetic pressure (gradient) can be included straightforwardly, as described in the first section and Appendix \ref{sec:Bfield}. } {We note that the formulae above are valid for the cHE stratification in general scenarios.}

\subsection{An isothermal example}
We extended the solution given by \citet{Zhang2021} by further including neutral and ionized helium (He and He$^+$), in addition to neutral and ionized hydrogen (H and H$^+$). Therefore, this example has two species and four fluids (not counting free electrons).  

 {More specifically}, we assumed an upper chromospheric condition adopted from a 2.5D radiative MHD (rMHD) model \citep{Martinez-Sykora_2020} using the \texttt{Bifrost} code \citep{Gudiksen2011}. The condition was found in regions where the ponderomotive force may produce chemical fractionation, associated with spicules or low-lying loops \citep{Sykora_2023}. Specifically, the temperature considered is $1.6\times10^4$ K and the  number densities of H, H$^+$, He, and He$^+$ were 10$^{9.8}$, 10$^{10.7}$, 10$^{9.7}$,  and 10$^{8.3}$ per cm$^{3}$, respectively. These number densities are a result of solving for nonequilibrium ionization (NEQ), namely, by solving the full time-dependent rate equations for hydrogen and helium \citep{Carlsson2002,Golding_2014}. When, instead, the rMHD model assumes statistical equilibrium (SE) ionization \citep{House_1964,mcwhirter1965,VORONOV1997}, by solving the rate equations without any $\partial N_i/\partial t$ term, the composition typically includes significantly more neutral fluids.  

 {Then}, for a nonmagnetic and isothermal ($T\equiv 1.6\times10^4$ K) case, we can easily obtain an analytical solution via
\begin{equation}
 p_{\text{T}}^{\text{g}}=p_{\text{T},z=0}^{\text{g}}\exp\left(-\frac{\overline{m}_{\text{T}}gz}{k_{\text{B}}T}\right).
\end{equation}
\noindent     {As a result, all fluids and species end up sharing the same pressure scale height}. We note that the  mean atomic mass is used for {balancing the plasma gas pressure gradient in Eq.~(\ref{eq:P_MFMS})}. When calculating the (mass) density of a specific fluid, we simply used its number density to multiply the real atomic mass of the fluid. In the present scenario, the ionization fractions do not change over height in the cHE stratification, and, thus, the (partial) number densities of all fluids can be easily calculated after the total gas pressure has been integrated. {To close the MFMS equations, we would need to have access to the temperature and partial densities, which can be used to calculate the partial pressures or internal energies following the equations of state.}

\begin{figure}[htpb]
\begin{center} 
    \includegraphics[width=0.5\textwidth]{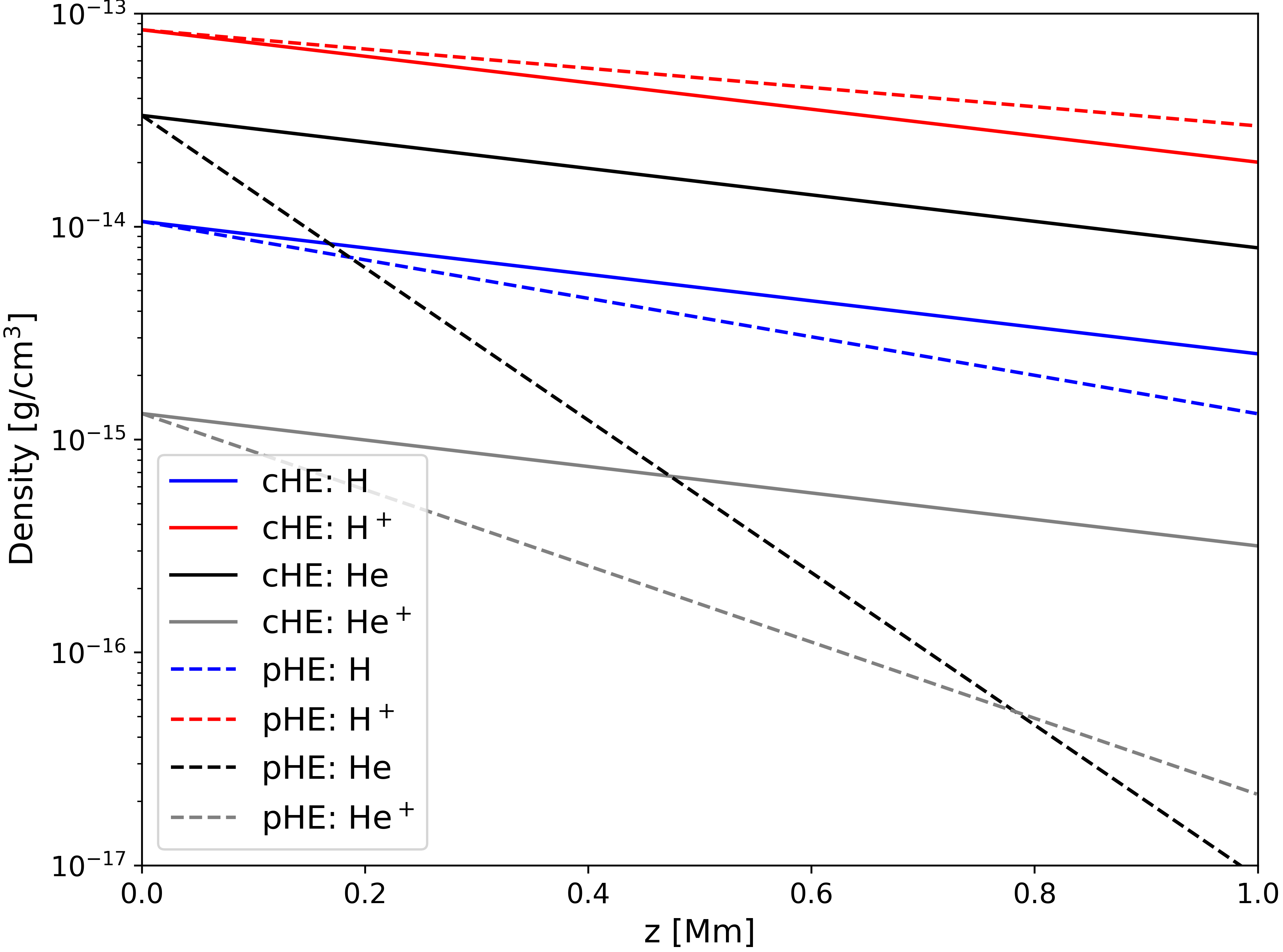}
\end{center}
\caption{Density stratifications of hydrogen-helium isothermal mixtures. The cHE stratification and the pHE stratification have the same ionization fractions at the reference height ($z=0$ Mm).}\label{fig:example}
\end{figure}

In Figure~\ref{fig:example}, we compare the pHE and cHE results given within a 1 Mm interval,\footnote{This interval is not suggested to be a realistic layer of the solar atmosphere.} using the aforementioned NEQ condition to provide the plasma quantities at $z=0$. The pHE solution results in exponentially changing ionization fractions,  {simply because of gravitational stratification}. In the cHE solution, each fluid is not in pure hydrostatic balance. Instead, gravity is balanced by the combination of the pressure gradient and the collisional forces, assuming that collisions are sufficiently strong. 

In a cHE stratification, the velocity of each fluid may not be zero, but their total velocity along the $z$ direction,
\begin{equation} \label{eq:totalV}
w_{\text{T}}=\frac{\sum_{i=1}^n \varrho_iw_i}{\sum_{i=1}^n \varrho_i},
\end{equation}
\noindent is expected to be within the integration error of Eq.~(\ref{eq:P_MFMS}).  {Specifically, when the temperature is constant, an analytical solution can be used and, thus, there is no error. In the following, we focus our discussion on general cases.}  A similar form of Eq.~(\ref{eq:totalV}) was introduced by \citet{Khomenko2014} as the velocity of the center of mass in the one-fluid model, which also explains the static assumption we used before any dynamic driving force was imposed. However, we did not include electrons when defining the total velocity, as the electron velocity was defined as a function of velocities of ions and the total current; namely, Eq.~(\ref{eq:ue}).

Given {that} the ionization fractions affect  {plasma properties}, it is incorrect to use the pHE to compute gravitational stratification, except in regions where collisional interactions and other mixing processes are negligible \citep{Hansteen_1997}, resulting in decoupled fluids.        {Having independent scale heights also makes the pHE stratification impractical in MFMS models for evaluating, for instance, the abundances of heavier elements}.  Overall, this example shows that it is crucial to construct cHE conditions with physical mixing processes, including, but not limited to, collisional interactions \citep{Hansteen_1994,Hansteen_1997}, particularly for MFMS modeling.

\section{General scenarios} \label{sec:general}
\subsection{A numerical integration routine}
In the previous section, we  constructed a coupled MFMS stratification that is in balance between gravity, pressure gradients, and collisional forces between multiple fluids and species.  
In more general scenarios, in which temperature is not constant, ionization fractions should vary       {even if SE is assumed}. Therefore, a numerical routine is needed       {to integrate Eq.~(\ref{eq:P_MFMS})}.  

Below, we assume that  {the temperature distribution is provided}. The ionization fractions can be calculated by assuming the SE condition under realistic solar atmospheric temperature distributions or provided by self-consistent rMHD simulations   with, for instance, NEQ ionization.  {Then, firstly,} we discretize Eq.~(\ref{eq:P_MFMS})  {in the 1D domain} and obtain a numerical  integration formula  {starting from the lower boundary of the domain}
\begin{equation}   \label{eq:routine}
p_{\text{T},j+1}^{\text{g}}=p_{\text{T},j}^{\text{g}}-\frac{p_{\text{T},j}^{\text{g}}\overline{m}_{\text{T},j}g_j}{k_{\text{B}}T_j}\Delta z_j, \quad j=0, ..., M,
\end{equation}
\noindent where $M$ is the number of intervals, with $\Delta z_j=z_{j+1}-z_j$, while  $\overline{m}_{\text{T},j}$  is calculated using Eq.~(\ref{eq:mt}),  {for which the partial number densities are needed},  and   $T_j$ is given and not changed by the integration routine. Also, $\overline{m}_{\text{T},j}$ and $T_j$ are considered to be constant  {within each integration interval}.      

At a discrete point, $j$,  {if} the  local ionization fraction is  {not provided in advance},  {it can be calculated using the given local temperature.} Except at $j=0$, where the boundary  {condition} is imposed,  {the local total gas pressure, $p_{\text{T},j}^{\text{g}}$, is provided by an integration step of $j-1$. The local total number density is calculated using the total gas pressure and the equivalent equation of state, Eq.~(\ref{eq:pT}). 
The ionization fraction and the total number density together provide} local partial number densities, with which  $\overline{m}_{\text{T},j}$  {can be calculated}.  {With $p_{\text{T},j}^{\text{g}}$ and $\overline{m}_{\text{T},j}$ provided,  $p_{\text{T},j+1}^{\text{g}}$ can be calculated.}

In practice, $M$ is preferably larger than the number of grid points used in numerical simulations. This is because the present integration routine is effectively a first-order left Riemann sum  {and, thus, the integration intervals need to be sufficiently small to reduce numerical errors}. The present solution is intuitively straightforward, as each integration step is an isothermal scenario introduced in the last section. However, the assumptions we used also meant that neither the hydrostatic equilibrium, nor the ionization equilibrium is strictly maintained between any pair of neighboring grid points due to numerical errors. Moreover, while collisional interactions are needed to balance gravity and pressure gradient, there must be drift velocities between fluids. Only the total velocity of all fluids was expected to be approximately equal to zero, as mentioned above. 

\subsection{Revisiting the SE assumption} \label{sec:re_SE}

\begin{figure}[htpb]
\begin{center} 
    \includegraphics[width=0.5\textwidth]{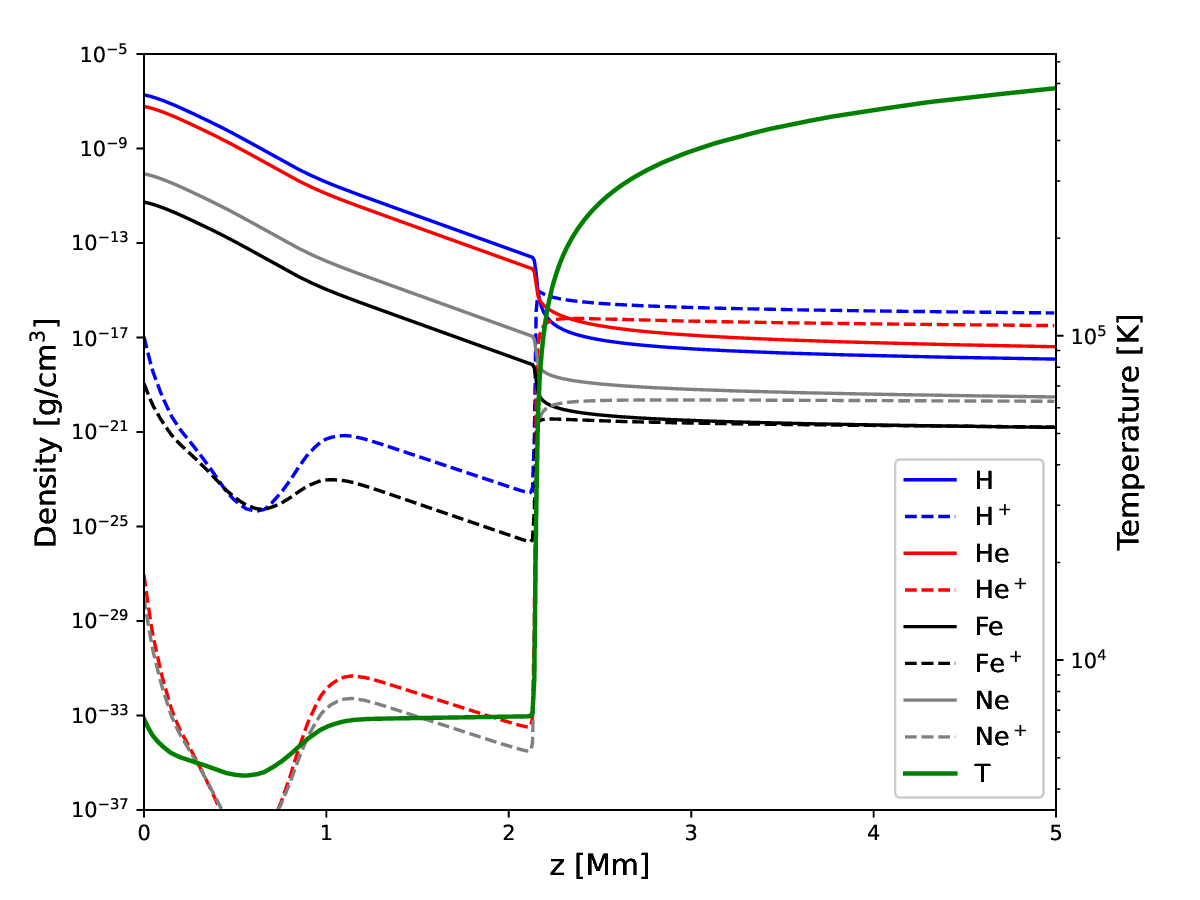}
\end{center}
\caption{The cHE stratification of a hydrogen-helium-iron-neon mixture under the model C7 temperature distribution.}\label{fig:SE}
\end{figure}

We calculated an SE stratification by first extracting the temperature distribution from the widely used model C7 \citep{Avrett2008}. The total number density at the bottom of the photosphere ($z=0$ Mm) was approximated using the hydrogen number density given by model C7, as ion and electron number densities are still small on the solar surface.  In addition to the four fluids included in Section \ref{sec:example}, we further include\footnote{      {There is effectively no limitation on the number of species that \texttt{Ebysus} can include, as long as relevant information (e.g., collisional cross sections) are available. Different species are included in this work simply to show the flexibility of the present approach.}} Ne, Ne$^+$, Fe,  and Fe$^+$. Then, SE ionization under the given temperature distribution is calculated to provide the number densities of these eight fluids, using the ionization and recombination rates of \citet{VORONOV1997} and \citet{mcwhirter1965}, with photospheric abundances of \citet{Asplund_2021} provided through the CHIANTI atomic database, Version 11.0 \citep{Dufresne_2024}. We note that we calculated only the cHE distributions here, because pHE would lead to significant deviations in the resulting ionization fractions, as explained in the previous section. The cHE stratification is shown in Figure~\ref{fig:SE}.

We find that the total ion density below the transition region (TR) is around ten orders of magnitude lower than the total neutral density and this is related to several factors. Firstly, the ionization state of hydrogen and helium is computed in SE, including collisional ionization balanced by radiative and three-body recombinations (for heavier ions, dielectronic recombinations would also be included), using rates from the CHIANTI package, while leaving out photoionization.  In addition, as mentioned, SE may underestimate ion fractions due to its ionization and recombination rates for some transitions. Thus, in the upper chromosphere, {nonequilibrium effects should be accounted for when describing the ionization and recombination processes}, while in the photosphere, solving the Saha equation \citep{D.Sc.1920,rutten2003} including all species is needed, because metals with their low FIP (first ionization potential) are the main donors of electrons. A more realistic calculation of SE ionization was presented by \citet{Gomez2024}, who used the MULTI code \citep{Carlsson_1986} to solve the radiative transfer equation together with the rate equations, thereby including photoionization in an appropriate fashion. However,       {how the ionization fractions are calculated does not affect the numerical integration}. Figure~\ref{fig:SE} shows that over the whole domain, the ionization fractions are in SE. Therefore, the present cHE stratification takes place both in hydrostatic equilibrium and SE.  

Moreover, we might also observe that within the temperature plateau of the chromosphere, the ionization fractions stay stable. This would indicate that when more realistic chromospheric dynamics are lacking, SE ionization cannot reproduce the increasing trend of ionization fractions in the upper chromosphere.  

\subsection{cHE stratification in NEQ} \label{sec:cHE_NEQ}

\begin{figure*}[htpb]
\begin{center}
    \includegraphics[width=0.75\textwidth]{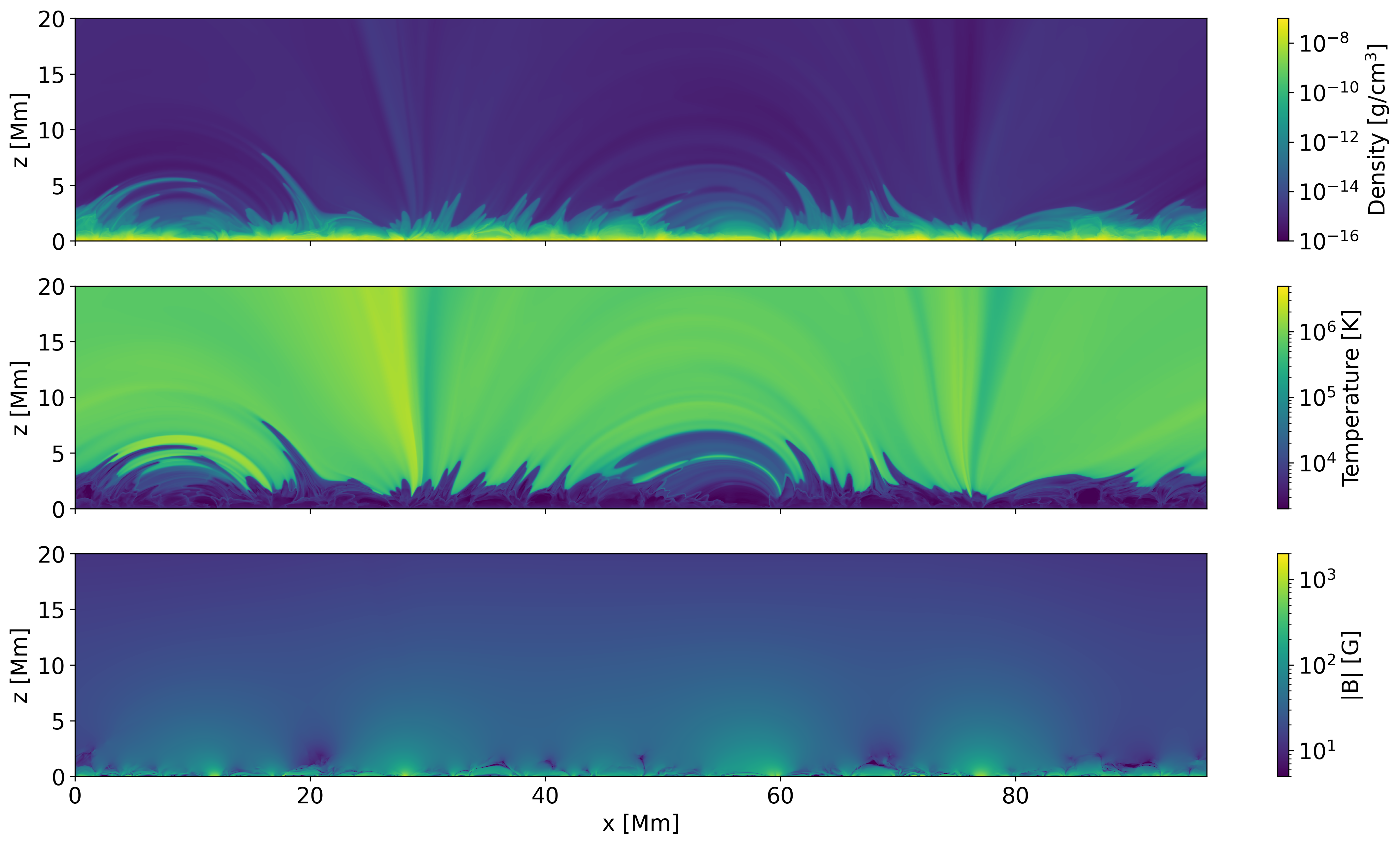}
\end{center}
\caption{A snapshot of a rMHD simulation with NEQ ionization.  Loops and spicules are observed in the snapshot. These structures are dynamic but also relatively steady compared to, for instance, high-frequency waves that may be responsible for local heating or the FIP effect.}\label{fig:2D}
\end{figure*}

The previous stratification is       {calculated using ionization fractions based on} the ionization and recombination rates in SE.       {However, NEQ ionization fractions can also be used to calculate the number densities needed for the numerical integration.} Below, we prove the flexibility of the present numerical integration routine, which is applicable as long as temperature and ionization fraction distributions are provided. 

\begin{figure*}[htpb]
\begin{center}
    \includegraphics[width=0.99\textwidth]{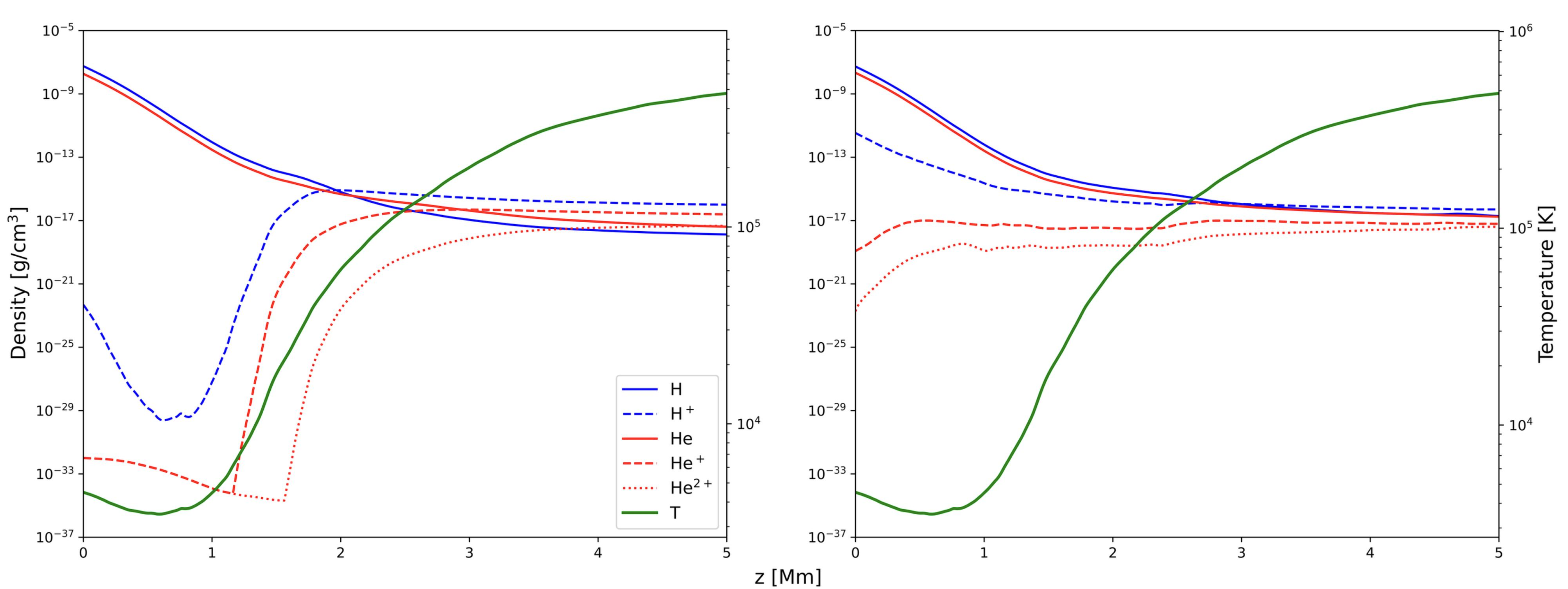}
\end{center}
\caption{Two cHE stratifications calculated based on the rMHD simulation snapshot. Left: Ionization fractions are calculated assuming SE and using the averaged total density and temperature from the rMHD model; right: NEQ ionization fractions are directly given as the averaged ionization fractions from the rMHD model. A small value is used as the floor when calculating ionization fractions in SE, which is a practical treatment for simulations, without affecting our conclusions.}\label{fig:cHE_Bifrost}
\end{figure*}

{As a schematic example}, we adopted a NEQ simulation snapshot from the 2.5D rMHD model \citep{Martinez-Sykora_2020}, as shown in Figure \ref{fig:2D}. In this model, we observe $\sim$50 Mm long loops, associated with the underlying magnetic field configuration, which has two main opposite polarities. Correspondingly, there are cold and dense features with chromospheric temperature along the loops, resembling spicules and jets \citep{Martinez-Sykora_2017}. More details of the rMHD simulation are not further discussed here. We calculated the horizontally-averaged density and temperature and then       {at first} used the method introduced in Section \ref{sec:re_SE} to obtain the SE ionization fractions. We note that the result here includes two-times ionized helium (He$^{2+}$) but not iron or neon, resulting in a two-species-five-fluid mixture. For comparison, we also extracted the averaged       {NEQ} ionization fractions from the 2.5D snapshot, replacing the SE ionization fractions, to perform the same integration routine. 
We obtained two stratifications shown in Figure~\ref{fig:cHE_Bifrost}, while omitting the contribution of magnetic pressure.  The inclusion of magnetic pressure is discussed in Appendix \ref{sec:Bfield}. We note that both stratifications are calculated using the same temperature distribution.

Due to the horizontal averaging, we obtain a relatively flat TR from the  rMHD model. Still, the aim is to show that a cHE stratification can be obtained to include the effects of NEQ ionization. Specifically, the NEQ stratification has higher ion populations in the chromosphere and photosphere,       {resulting from}  the long timescales of the ionization and recombination processes, as well as their interplay with chromospheric dynamics \citep{Carlsson1992,Carlsson2002,Leenaarts2007,Wedemeyer_B_hm_2011,Golding_2014}. Therefore, the present numerical integration routine provides flexibility for specifically studying atmospheric dynamics, while considering the consequence of NEQ ionization. In contrast, a pHE stratification cannot recover such effects, as shown in Figure \ref{fig:example}.

\section{MFMS equations and numerical methods} \label{sec:ebysus}

In the following, we ran 1D MFMS numerical simulations that use the stratifications considered above as initial conditions. We used the numerical code \texttt{Ebysus} \citep{Mart_nez_Sykora_2020,Sykora_2023,Wargnier_2023,Wargnier_2025}.
 Compared to various multifluid codes \citep{Laguna2016,PopescuBraileanu2019,Braileanu_2022,Snow_2023}, an important feature of \texttt{Ebysus} is that it can solve the MHD equations separately for each of the desired number of excited levels,
ionization stages, and species, although in this work, we solve the collective motion of all the excited levels of each neutral species for simplicity. 
 \texttt{Ebysus} is modular and allows for the inclusion of ionization and excitation in NEQ, collisions between different fluids, thermal conduction, radiative losses, and the Hall term. Details of the derivation of the MFMS formulae have been introduced by \citet{Khomenko2014, Ballester2018}. For the purposes of this work, we neglected radiation, thermal conduction, and the Hall term. We also considered only 1D.

In \texttt{Ebysus}, the momentum equation of electrons is used to calculate the electric field, ignoring the inertia of the electrons and their time variation.  The electron number density is not an independent variable but is determined by the number densities and ionization states of all the ions, namely, 
\begin{equation}  
N_{\text{e}}=\sum\nolimits_{aI}N_{aI}Z_{aI},
\end{equation}
\noindent where $Z$ is the ionized state, $a$ indicates the identity of the chemical species, and $I$ denotes the ionization states, with $I=0$ denoting neutrals and $I>0$ ions. In addition, as the quasi-neutrality approximation is assumed, the electron velocity is expressed as a function of the hydrodynamic velocities of ions and the total current, $\mathbf{J}=\nabla\times\mathbf{B}/\mu_0$, namely,
\begin{equation}  \label{eq:ue}
\mathbf{u}_{\text{e}}=\sum\nolimits_{a{I}}\frac{N_{a{I}}q_{a{I}}\mathbf{u}_{a{I}}}{N_{\text{e}} q_{\text{e}}}-\frac{\mathbf{J}}{N_{\text{e}} q_{\text{e}}},
\end{equation}
\noindent where $q_{a{I}}$ is the ion charge. We note that even when using fewer species, similar formulae were used by \citet{Martinez-Gomez2016}.  

More importantly, \texttt{Ebysus}  solves a separated energy equation of electrons along with the energy equations of ions and neutral fluids. Consequently, the contribution of electron pressure (gradient) is not directly included in the momentum and energy equations of ions. Meanwhile, the electron pressure contributes to the electric field, which is then included in the momentum equations of ions. Thus, electron pressure also contributes to the induction equation, which reads 
\begin{equation} \label{eq:dBdt}
\frac{\text{d} \mathbf{B}}{\text{d} t}=\nabla\times\mathbf{E}=\nabla\times\left(\mathbf{u}_{\text{e}}\times\mathbf{B}-\frac{\nabla p_{\text{e}}}{N_{\text{e}}q_{\text{e}}}-\frac{\sum\nolimits_{aI}\mathbf{R}_{\text{e}}^{eaI}}{N_{\text{e}}q_{\text{e}}}\right),
\end{equation}
\noindent where $\mathbf{R}_{\text{e}}^{eaI}$ denotes the collisional interactions between electrons and other fluids \citep{Mart_nez_Sykora_2020}. Therefore, the electron pressure gradient (thus the Biermann battery term, namely, the second term on the right-hand side of Eq.~(\ref{eq:dBdt})) is necessary for \texttt{Ebysus} to account for the contribution of electron pressure in gravitationally stratified atmospheric models.

To spatially discretize the governing equations, \texttt{Ebysus} uses a sixth-order finite-difference scheme with hyper-diffusivity. A fifth-order interpolation is used to interpolate variables in the staggered grid system. To advance in time, we used the modified explicit third-order predictor-corrector Hyman method \citep{Hyman1979}. These numerical approaches are inherited from the \texttt{Bifrost} code \citep{Gudiksen2011}. 
Recently, we included the second-order partitioned implicit-explicit orthogonal Runge-Kutta (\texttt{PIROCK}) method in \texttt{Ebysus} \citep{ABDULLE2013869,Wargnier_2025}, allowing for advancements over time by combining efficient explicit stabilized and implicit integration techniques, while employing variable time-stepping with error control. The \texttt{PIROCK} method was also tested in this work, further supporting the present results. 

\section{Remnant disturbances and static decoupling} \label{sec:disturbances}

\subsection{Initial conditions}
A cHE stratification obtained using the numerical integration routine is approximately in equilibrium. Therefore, there could be remnant disturbances when a simulation uses the cHE stratification as the initial condition. More importantly, according to our assumption, there must be drift velocities between fluids. To evaluate these effects, we simulated disturbances developing in initially quiet stratifications, without imposing any external dynamic driving force. Specifically, we constructed a cHE stratification using the model C7 temperature distribution, yet including only\footnote{The model C7 provides discrete data points. Therefore, the temperature distribution is interpolated using the \texttt{interp1d} function from the \texttt{NumPy}  library \citep{harris2020array} with the \texttt{quadratic} interpolation method. Using the \texttt{linear} method would have caused more significant errors in drift velocities and, eventually, in the total velocity.} 2.2 to 3.2 Mm and hydrogen and helium (no He$^{2+}$). Therefore, the domain does not include the non-smooth transition from the upper chromosphere to the TR at around 2.14 Mm in model C7.  We included only the smooth (yet steep) variation above the transition, where the pressure gradient is large. 

\begin{figure*}[htpb]
\begin{center} 
    \includegraphics[width=0.99\textwidth]{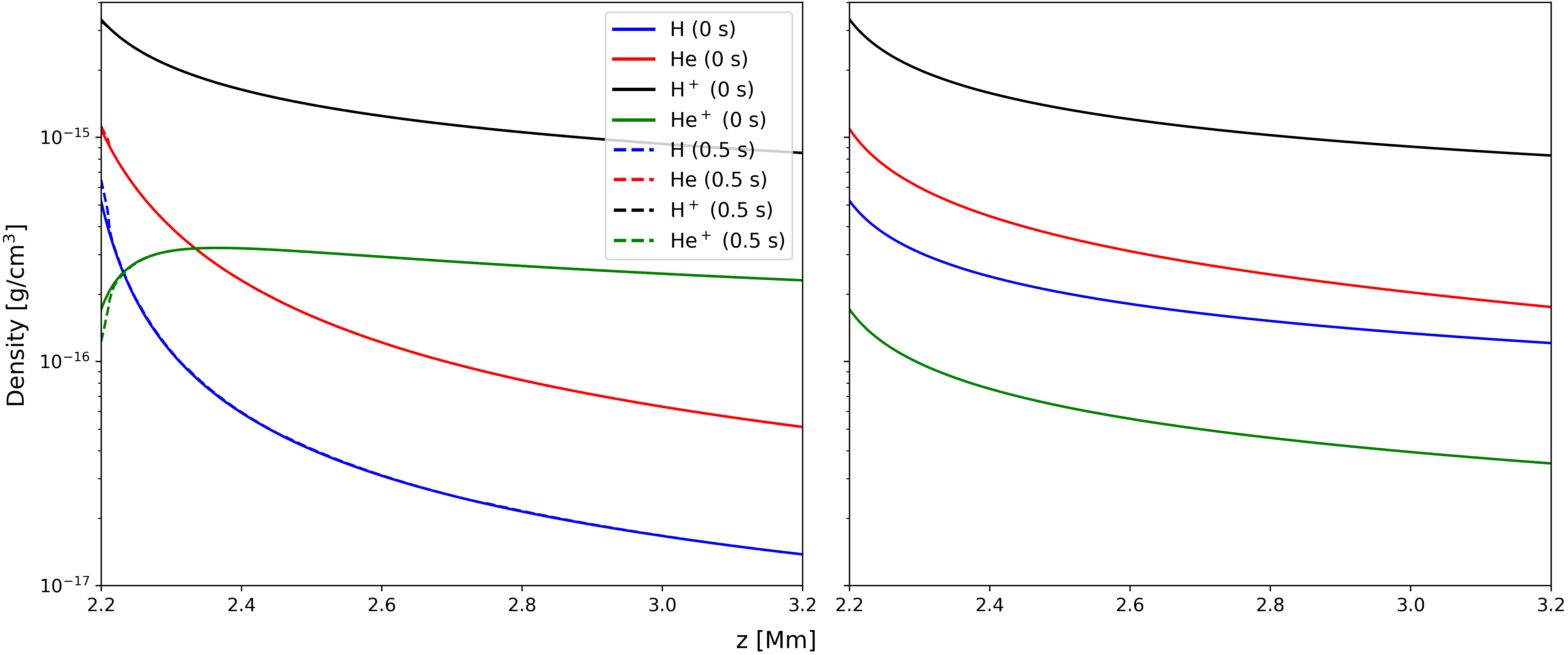}
\end{center}
\caption{Left: Density distributions at $t=0$ s and $t=0.5$ s when using the cHE stratification as the initial condition for a simulation with NEQ ionization and recombination, but without any dynamic driving force. Right: pHE stratification (viz. at $t=0$ s). Using the pHE stratification as the initial condition leads to numerical instability in the simulation with ionization/recombination and, thus, the result is not included. }\label{fig:cHE_pHE_1Mm}
\end{figure*}

\begin{figure}[htpb]
\begin{center} 
    \includegraphics[width=0.5\textwidth]{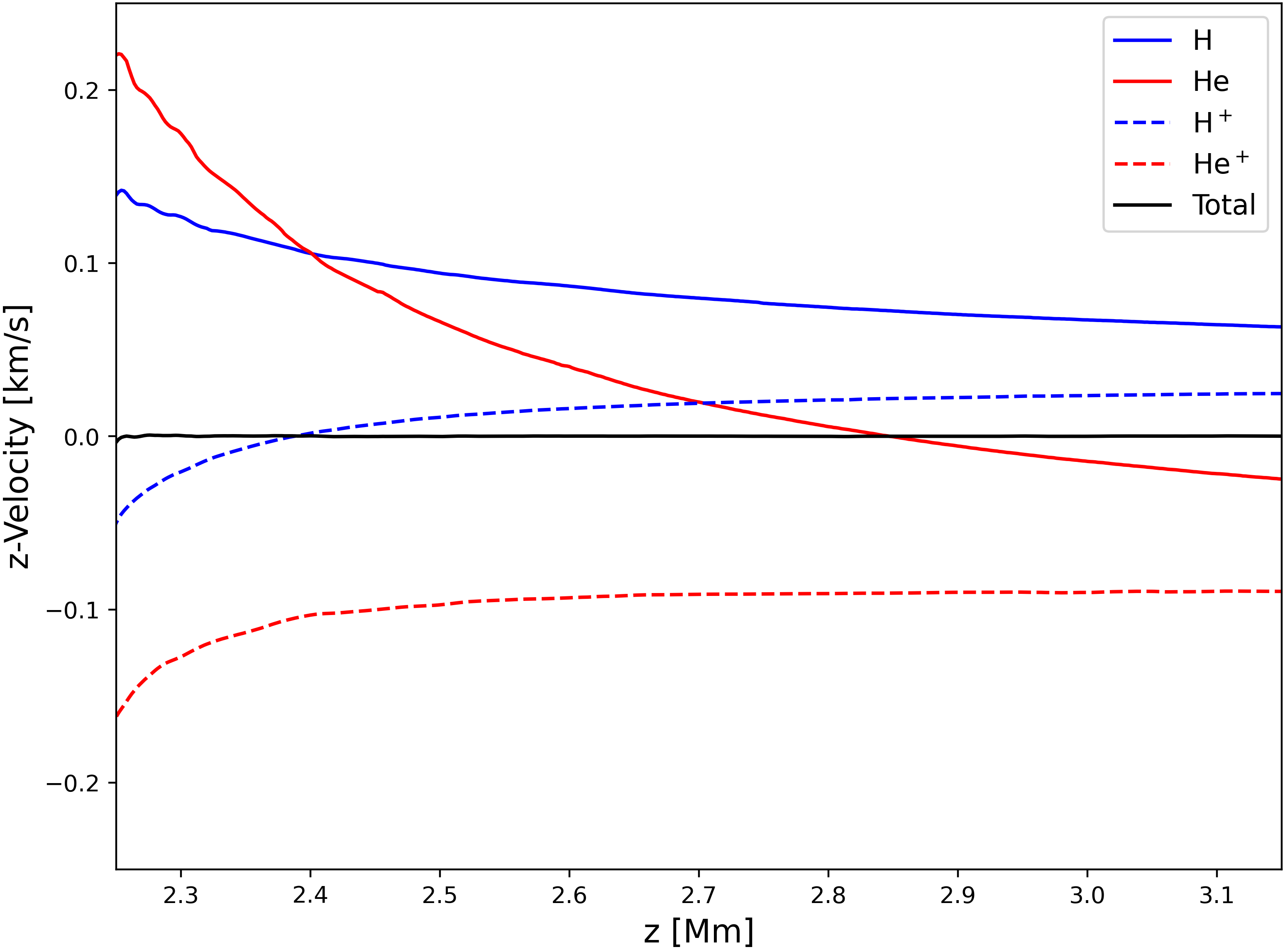}
\end{center}
\caption{Longitudinal velocities of the cHE + NEQ simulation shown in Figure \ref{fig:cHE_pHE_1Mm} at $t\approx 0.5$ s. The total velocity and the velocities of all fluids are shown. The initial velocities are all zero.}\label{fig:cHE_pHE_vz}
\end{figure}

\begin{figure}[htpb]
\begin{center} 
    \includegraphics[width=0.5\textwidth]{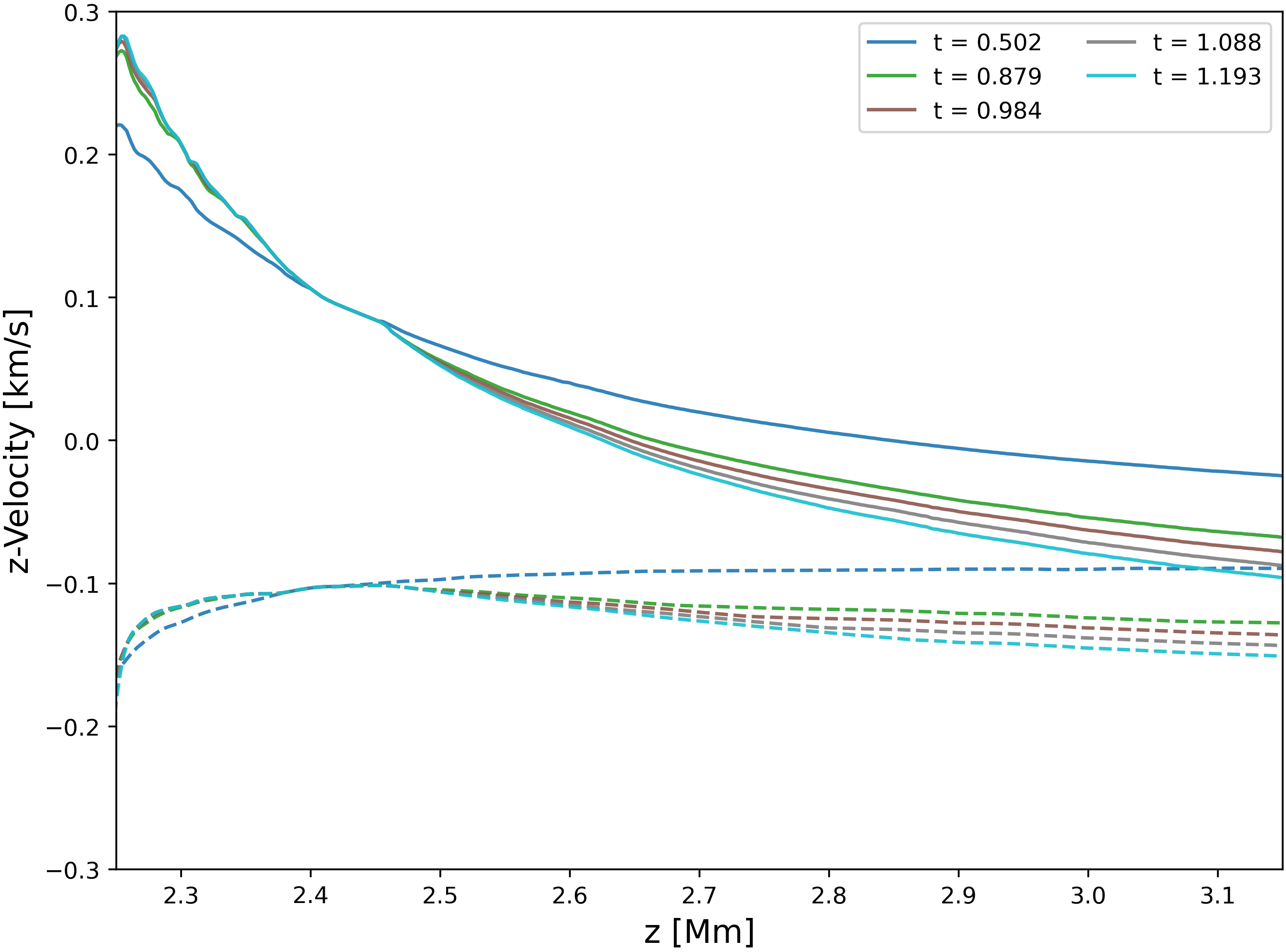}
\end{center}
\caption{{Temporal variations of the velocity distributions of neutral helium  (solid lines) and ionized helium (dashed lines) in the cHE + NEQ simulation.}}\label{fig:cHE_He_vz}
\end{figure}

The resulting initial cHE stratification (Figure~\ref{fig:cHE_pHE_1Mm}, left panel in solid lines) is similar to the result shown in Figure~\ref{fig:SE} for the selected range. 
 Across the whole domain of the cHE stratification, plasma is approximately in SE and, thus, higher up in the domain, the helium ionization fraction continues to increase with increasing temperature. Correspondingly, the fractions of neutrals gradually decrease. For comparison, the initial pHE stratification constructed using the same plasma conditions at $z=2.2$ Mm as the cHE stratification is shown with solid lines in Figure~\ref{fig:cHE_pHE_1Mm}, right panel. In the initial pHE stratification, the ionization fractions do not change with the (rapidly) increasing temperature, and thus the plasma is not in SE. The initial condition is imposed on 500 evenly distributed grid points that discretize the domain. The mesh is sufficiently fine for the present simulations and, {thus,} we can avoid the effect of numerical diffusion.

\subsection{SE evolution}

 Using the given initial conditions, we run simulations without any additional dynamic driving force. The simulations use physically calculated collisional frequencies \citep{Wargnier_2022}, and include the NEQ ionization and recombination processes. We compare the temporal variations of densities in the simulation using the cHE stratification as the initial condition, as shown with dashed lines in Figure \ref{fig:cHE_pHE_1Mm} (left panel). The density distributions change little over the simulated time, except near the lower boundary, where more accurate spatial discretization is usually needed \citep{Krause2019,Wei_2023}. Therefore, we may claim that the initial condition is sufficiently close to SE in the whole domain. However, the simulation starting with the pHE stratification develops numerical instability as a result of the rapid ionization caused by the initial nonequilibrium and, thus, the corresponding result is not further discussed. 

Although the total velocity in the cHE + NEQ simulation is not exactly zero, only small disturbances slowly develop over time, {most significantly} in the lower part of the domain (Figure \ref{fig:cHE_pHE_vz}).  The decoupling in the cHE result causes collisional forces. In the meantime, the ionization and recombination processes exchange momentum between ions and neutrals, reducing the decoupling.       {The velocity drift of neutral helium is the most significant, but its increase is evidently slowing down over time, as shown in Figure \ref{fig:cHE_He_vz}. In particular, as we can see in the lower part of the domain, the velocities soon ($\sim$1 s) stop changing, while in the upper part, the relative drifts between ionized and neutral fluids also become stable. In the upper part, both neutral and ionized helium move downwards, whereas hydrogen moves upwards (not shown) and, thus, the total velocity remains around zero (Figure \ref{fig:cHE_pHE_vz}). In this specific case, the smaller pressure gradients higher up cause less  decoupling, but the collisional interactions are also weaker to stabilize the decoupling.} When ionization and recombination are not included, the decoupling changes over time at a faster rate, until the collisional forces are able to balance the gravity and the pressure gradients, in which case the maximum decoupling between the neutral helium and the total velocity reaches above 2 km s$^{-1}$ (not shown). The drift velocities that occur in the cHE stratification are physically important. In particular, such drift velocities make up an important difference between single-fluid and multifluid models \citep{mguez2025}.  

These results suggest that the pHE stratification is not viable for simulations that include ionization and recombination, as numerical instability may occur. Even if ionization and recombination are not activated, the ionization fractions are important for correctly describing the collisional interactions. For example, in Figure \ref{fig:cHE_pHE_1Mm}, the density of the neutral hydrogen in the pHE stratification can be around an order of magnitude higher than that of the neutral hydrogen in the cHE stratification and this difference certainly affects collisional frequencies.

Moreover, the cHE model provides a steady solution with chemical fractionation for a gravitational settling of the solar atmosphere, accounting for the effects of ionization and recombination. We note that in the solar atmosphere, dynamic phenomena such as MHD waves may also be responsible for chemical fractionation \citep{Laming_2004,Laming2015}. In addition, the current cHE stratification can be used to numerically model the dynamic effects.  

\section{The ponderomotive force and the resulting motions} \label{sec:waves}
Since significant decoupling appears in the cHE stratification when the pressure gradient is large, it is interesting to investigate how this "static" decoupling would contribute to the MFMS dynamics. In this study, we ran simulations of a monochromatic Alfv\'en wave, using the cHE and pHE stratifications introduced in the previous section. It has been suggested that Alfv\'en waves existing in the low solar atmosphere could play a role \citep{Pontieu2007,McIntosh_2011,Chae_2021}, particularly with respect to coronal heating \citep{VanDoorsselaere2020}, the FIP effect \citep{Laming_2004,Laming2015}, and other phenomena. Here, we are interested in exploring the fact that the ponderomotive force induced by Alfv\'en waves drives plasma motions along the magnetic field \citep{Lundin2006}. Such motions can be simulated and used as an example to compare the effects of using different stratifications.

We imposed an initially constant vertical magnetic field of $B_z=10$ G on the stratifications discussed. The Alfv\'en wave is driven by periodically changing $B_x$ at $z=2.2$ Mm, similar to what was done in the study carried out by \citet{Sykora_2023}. We note that such a driver might not be suitable in the weakly ionized photosphere \citep{Cally_2023}, but it does work under the present atmospheric condition. The frequency of the driver is 1 Hz, and the amplitude is 1 G. The resulting amplitude of the velocity along the $x$ direction is $\sim 40$ km s$^{-1}$. Based on this basic setting, five simulations are performed, as listed in Table \ref{tab:cases}. We set our focus on comparing the differences between using cHE and pHE stratifications, with or without the NEQ ionization (simulations (c)-(e)). Because of the numerical instability mentioned in the previous section, the pHE stratification is only discussed without including ionization and recombination in the simulation (simulation (d)). We also included two simulations that use nonphysical collisional frequencies to examine the weakly and strongly collisional scenarios (simulations (a) \& (b)). These two nonphysical frequencies are significantly lower and higher than the imposed wave frequency, respectively. The former should evolve as separated fluids, while the latter should recover the single fluid scenario. 

\begin{table}[ht]
\centering
\caption{Summary of the simulations.}
\begin{tabular}{c c c c c}
\hline
& Stratification &  Ionization &   Collisional frequency \\
\hline
(a) & cHE & -- &  0.05 Hz  \\
(b) & cHE & -- & 100 Hz  \\
(c) & cHE & -- & physical   \\
(d) & pHE & -- & physical  \\
(e) & cHE & NEQ & physical  \\ 
\hline
\end{tabular}
\label{tab:cases}
\end{table}

\begin{figure*}[htpb]
\begin{center} 
    \includegraphics[width=0.9\textwidth]{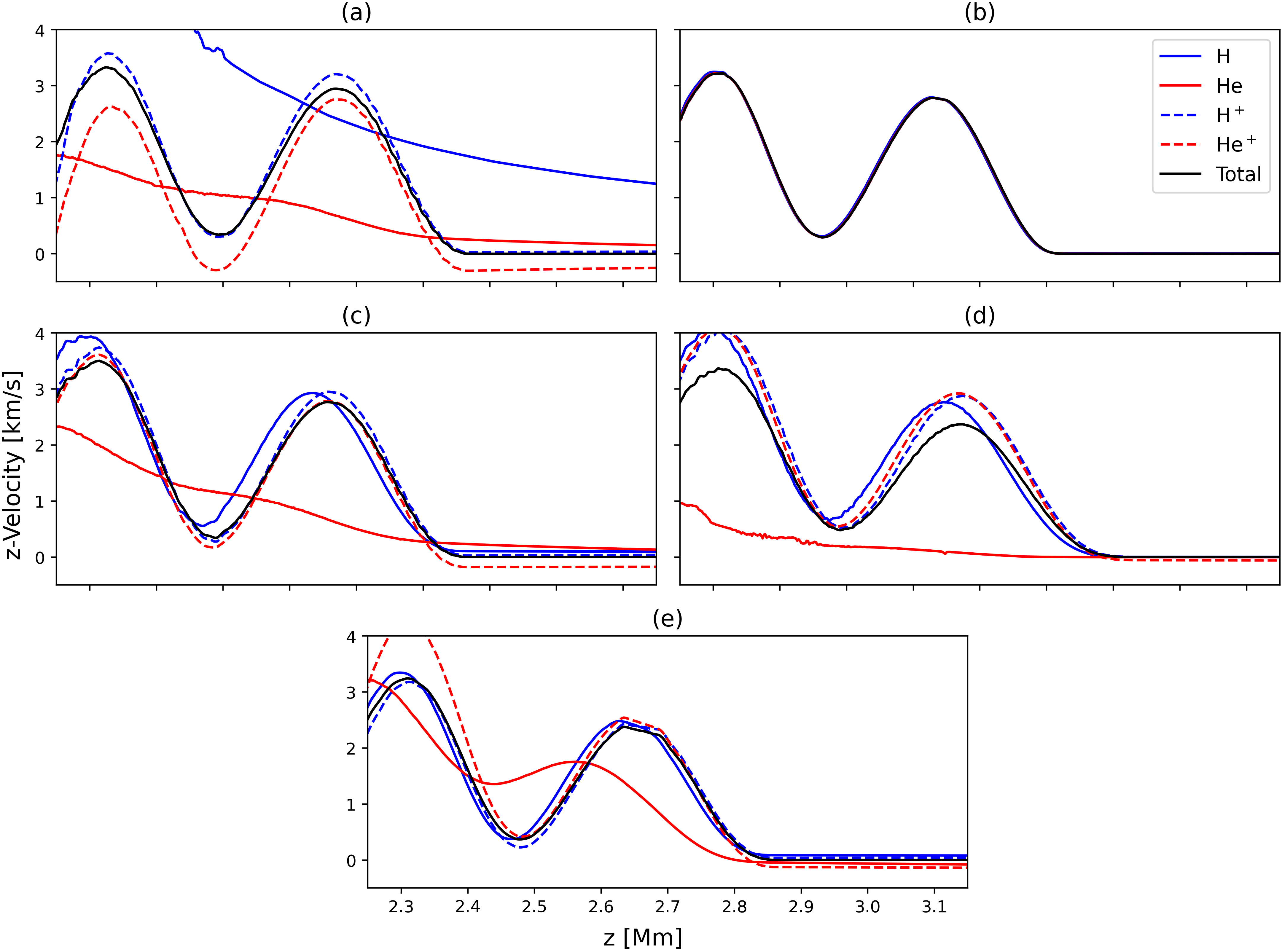}
\end{center}
\caption{Longitudinal velocities at $t=1$ s. All settings are listed in Table \ref{tab:cases}. Except simulation (d), which uses the pHE stratification,  all the others use the cHE stratification as the initial condition. Simulations (a) and (b) use assigned constant collisional frequencies. More specifically, (a) is weakly collisional and (b) is strongly collisional. All the other simulations use physical collisional frequencies calculated by \texttt{Ebysus}. Only simulation (e) includes the NEQ ionization/recombination.  }\label{fig:waves}
\end{figure*}

The effects of the numerical disturbances appear in the $z$ direction (i.e., the direction of the magnetic field). The Alfv\'en wave-induced ponderomotive force       {relates to $-{\partial E^2_{\perp}}/{\partial z}$, in which $E_{\perp}$ is the electric field perpendicular
to the magnetic field \citep{Lundin2006,Sykora_2023}. Therefore, upward motions (positive longitudinal velocity) are driven by the ponderomotive force, as shown} in Figure \ref{fig:waves}. 
When it is weakly collisional (panel a), the neutral fluids are decoupled from the ionized fluids, whose motions are driven by the Alfv\'en wave. Each neutral fluid also moves independently because of the imbalance between gravity and the respective plasma gas pressure, while neither collisions with ionized fluids nor magnetic pressure affect neutral fluids. In contrast, all fluids in the strongly collisional simulation are coupled, behaving as a single-fluid as expected (panel b). More importantly, we can observe that in the strongly collisional simulation, the total velocity is zero before the wave arrives, showing that the cHE stratification is indeed in hydrostatic equilibrium, providing another way to validate the MFMS code; namely once it is collisionally coupled, the plasma behaves as a single fluid. 

Simulations using physical collisional frequencies \citep{Wargnier_2022} show significant decoupling of neutral helium (panels c-e). A similar result was also obtained by \citet{Martinez-Gomez2018}, although only initially homogeneous media were considered. \citet{Wargnier_2023} also found that neutral helium is more significantly decoupled from other fluids during upper-chromospheric magnetic reconnection. Thus, including helium may be important for modeling the dynamics across the TR. Moreover, when using cHE stratification (panels c and e), neutral helium has a higher longitudinal velocity than pHE (panel d), due to the contribution of the pressure gradient, which is more obvious in Figure \ref{fig:cHE_pHE_vz}. The ionization process strengthens the interactions between ionized and neutral fluids, thereby increasing the coupling in the simulation (e). Nevertheless, in general, neutral helium has a smaller longitudinal velocity than the other fluids.

The pHE stratification is not further discussed with ionization and recombination due to the numerical instability. We should, however, note that the ionization and recombination timescales are long in the chromosphere \citep{Carlsson2002} and, thus, the changes in density and temperature would be less drastic when the pHE stratification is used under chromospheric conditions. However, this also means that the ionization fractions (and thus the collisional interactions) in the corresponding simulations would be largely dependent on the initial stratification. We have shown in Section \ref{sec:example} that heavier elements, whose scale heights are different from that of (ionized or neutral) hydrogen, cannot be properly recovered by the pHE stratification, even if the collisional interactions (or other mixing processes) are sufficient to couple the heavy fluids.

\section{Conclusions and outlook}

The primary goal of this work is to provide gravitationally stratified initial and background fields for MFMS modeling. This goal was achieved using a simple numerical integration routine based on relaxed equilibrium assumptions. Specifically, as an initial static condition is given, we can assume that all fluids can be coupled by the collisional interactions, which might then contribute to the hydrostatic equilibrium (viz., the cHE stratification) and allow us to account for any expected ionization fractions, particularly with respect to the SE ionization fractions. Without such initial fields, it is difficult to include both gravitational stratification plus ionization and recombination simultaneously, since instability or disturbances caused by initial {imbalances} in the pHE stratification can lead to significant errors and even sabotage numerical modeling.  The present work is even more critical when more species are needed, whose pressure scale heights are vastly different, in which case the purely hydrostatic stratification (viz., the pHE stratification) is far from being in ionization equilibrium and heavy elements effectively disappear. The cHE stratification provides a gravitational settling for chemical fractionation, whereby heavy elements are (partly) supported by collisional interactions, with no other dynamic driving mechanisms included.

The present cHE stratification also clearly shows the drift velocities accompanied by strong pressure gradients, which are more prominent in the TR. Such drift velocities are deemed physical, as the pressure scale heights of different fluids are more significantly different in this region. In contrast, the pHE stratification does not recover such effects, regardless of the pressure gradients and, thus, there is no collisional interaction between fluids, which is a strong hypothesis.

Eventually, the present cHE provides a gravitationally stratified "static" atmospheric model that can be used to more specifically and accurately study waves, reconnection, and so on, potentially improving previous MFMS models further \citep{Martinez-Gomez2016,Martinez-Gomez2017,Martinez-Gomez2018,Mart_nez_Sykora_2020,Wargnier_2023,Sykora_2023}. We note that multifluid effects have been discussed with regard to various topics and including (only) hydrogen and helium suffices in many cases. However, numerical models of the FIP effect, particularly, need to explicitly include heavier elements, which has been challenging. The present work shows that, for example, the ponderomotive force and the resulting motions can be modeled using the cHE stratification, which allows gravitational stratification and NEQ ionization to be incorporated into the previously semi-homogeneous MFMS model \citep{Sykora_2023}.  

In addition, we artificially increased (or decreased) the collision rates, producing strongly (or weakly) coupled models behaving as a single fluid (or independent fluids). This parametric study briefly illustrates the importance of accurately describing fluid interactions, which can substantially alter plasma properties \citep{Hansteen_1997}. Moreover, the result further shows that the \texttt{Ebysus} code is capable of coupling all the fluids when interactions are sufficiently strong.

 In this work, we only discuss (quasi-)1D scenarios, in the case of which it is straightforward to account for the contribution of the magnetic field. Similarly, the cHE stratification can be directly used under multidimensional force-free magnetic fields. The extension to multidimensional non-force-free scenarios will be addressed in future works.
 
\begin{acknowledgements}
The authors thank Nicolas Poirier, Mats Carlsson, Boris Gudiksen, and Ben Snow for the helpful discussion and Mikolaj Szydlarski for technical support.
This research has been supported by the Research Council of Norway through its Centres of Excellence scheme, project number 262622, and by computational resources provided by Sigma2, the National Infrastructure for High Performance Computing and Data Storage in Norway.
JMS acknowledges support by NASA contracts NNG09FA40C (IRIS) and 80GSFC21C0011 (MUSE),  NASA grant 80NSSC26K0018, NSF grants AGS2532363 and AGS2532187.
This work used the CHIANTI atomic database.  CHIANTI is a collaborative project involving George Mason University, the University of Michigan (USA), the University of Cambridge (UK), and NASA Goddard Space Flight Center (USA). Figures in this work
were produced using the Python package \texttt{Matplotlib} \citep{Hunter_2007}. Data processing was performed using the \texttt{NumPy} library \citep{harris2020array} and in-house packages \texttt{EbysusRunTools\_py}, \texttt{atom\_py} and \texttt{Helita}.

\end{acknowledgements}
 
\bibliographystyle{aa}
\bibliography{refs}

\begin{appendix}

\section{Effects of magnetic field} \label{sec:Bfield}

\begin{figure}[htpb]
\begin{center} 
    \includegraphics[width=0.5\textwidth]{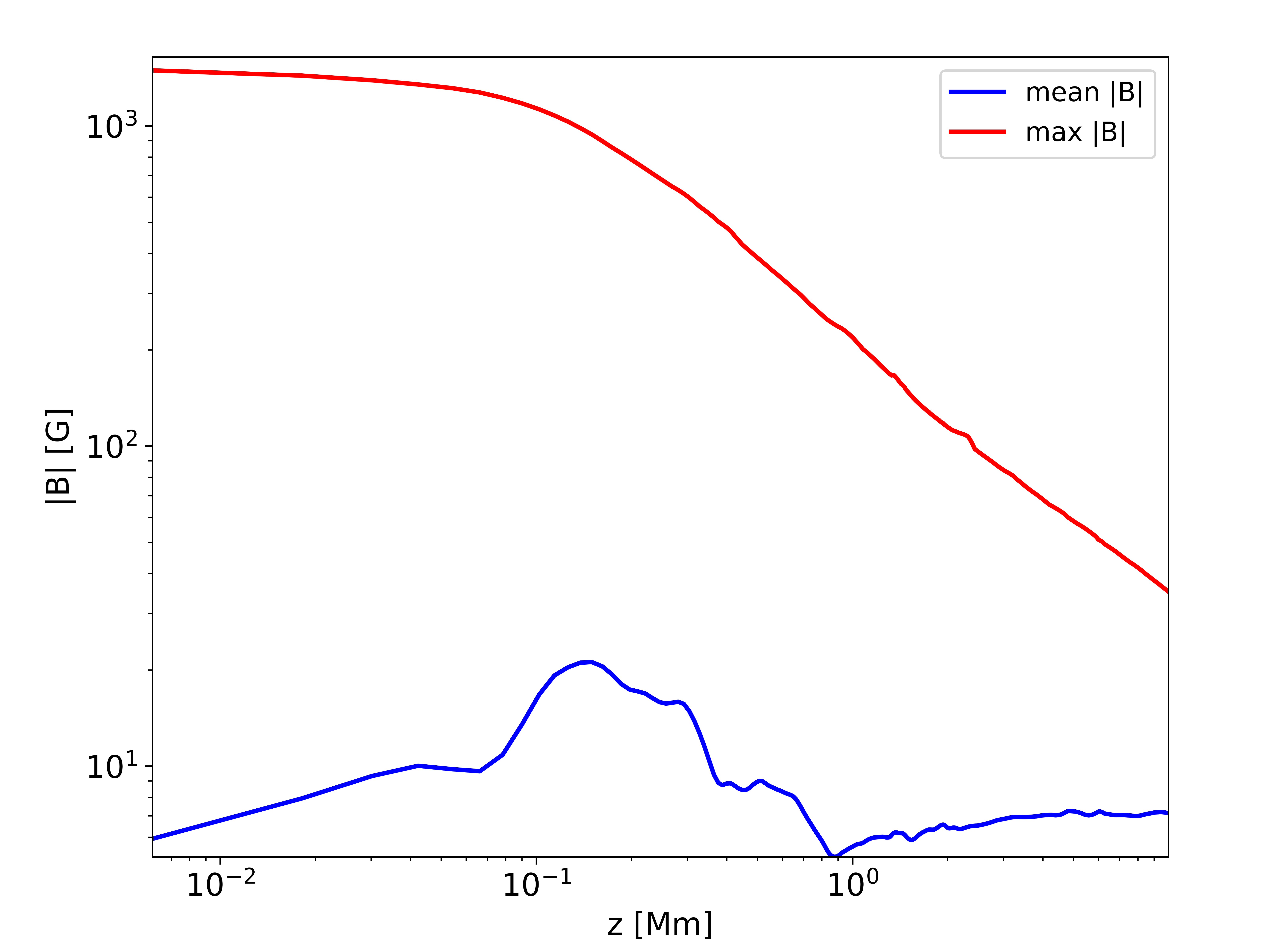}
\end{center}
\caption{The maximum and averaged $|\mathbf{B}|$ at each height. The maximum value shows an inverse power law above $\sim$ 0.1 Mm. The averaged value first drastically increases and then gradually decreases, and exhibits slower variation above $\sim$1 Mm.}\label{fig:BField}
\end{figure}

The magnetic field is a driving force of the complex dynamics in the solar atmosphere. In particular, locally concentrated magnetic structures in the photosphere may undergo rapid expansion in the low solar atmosphere \citep{Gary2001,Solanki_2006}, resulting in large magnetic pressure gradients that cannot be overlooked when considering gravitational stratification. 

Therefore, following the results in Section \ref{sec:cHE_NEQ}, we discuss the effects of adding a magnetic field. The magnetic field in this 2D model is locally concentrated in the photosphere and shows a significant expansion higher up. Consequently, the unsigned mean
magnetic field strength decreases over height (Figure~\ref{fig:2D}, bottom). Observing Figure~\ref{fig:BField}, a power law can be seen from the variation of the maximum value at each height, but it is not representative of the whole magnetic field. The averaged value also cannot represent the derivative of the magnetic field.  Nevertheless, we do not  focus on such details, and the results justify using an expansion assumption to approximate the effect of magnetic pressure.

\begin{figure}[htpb]
\begin{center} 
    \includegraphics[width=0.5\textwidth]{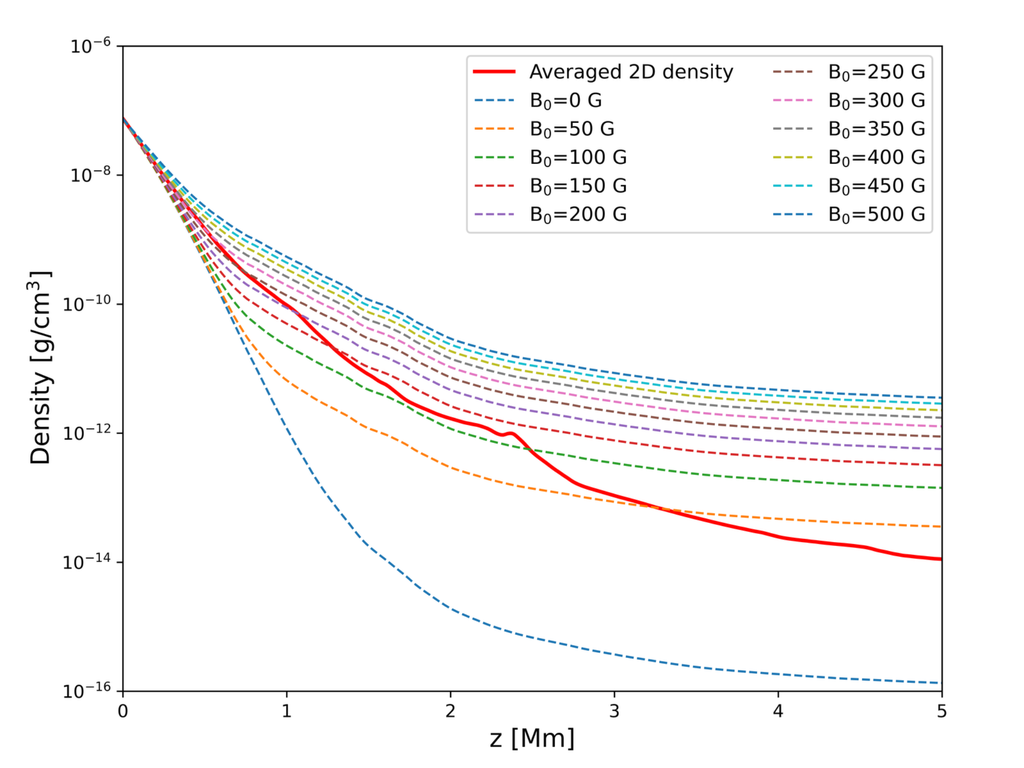}
\end{center}
\caption{cHE stratifications calculated with the given magnetic pressure distributions,  $r_0=1$ Mm, and SE ionization fractions.      The total density when $B_0=0$ G is the same as the SE result in Figure \ref{fig:cHE_Bifrost}.}\label{fig:total_density}
\end{figure}

\begin{figure}[htpb]
\begin{center} 
    \includegraphics[width=0.5\textwidth]{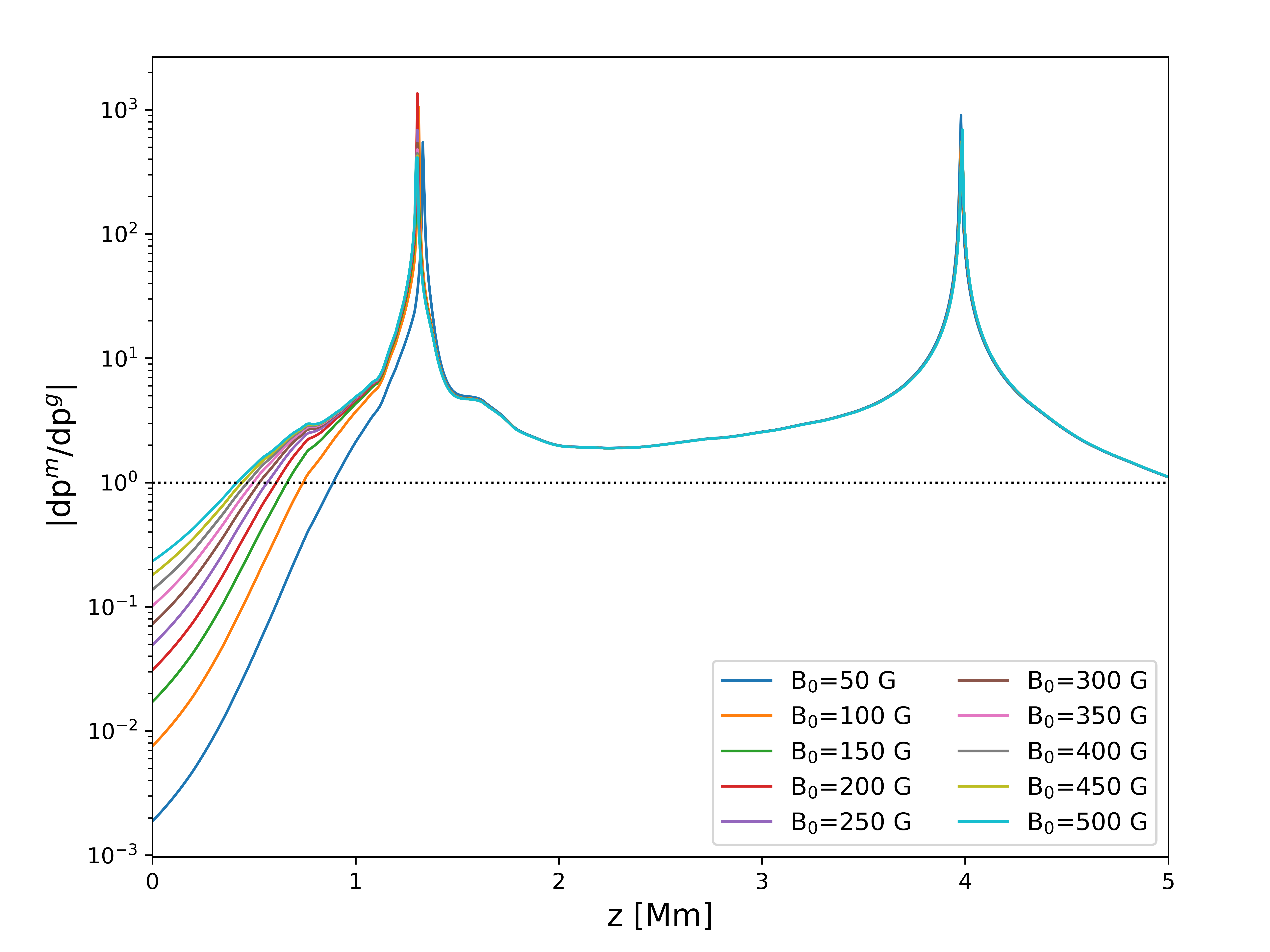}
\end{center}
\caption{Ratios of the magnetic pressure gradient to the plasma gas pressure gradient. The peaks appear because the plasma gas pressure gradient changes sign. Above the photosphere, the magnetic pressure gradient becomes dominant.}\label{fig:ratio_comparison}
\end{figure}

Assuming that the (3D) spatial expansion is the only cause reducing the magnetic field strength, we  use 
\begin{equation} \label{eq:Bz}
p^{\text{m}}(z)=0.5\left[\frac{|\mathbf{B}_0|}{(z+r_0)^2/r_0^2}\right]^2,
\end{equation}
\noindent where $\mathbf{B}_0$ and $r_0$ are free parameters, to describe the magnetic pressure. With Eq.~(\ref{eq:Bz}), a smaller $r_0$ means that the magnetic field is more concentrated in the photosphere. The pressure gradient can be analytically calculated based on Eq.~(\ref{eq:Bz}), and directly added to the numerical integration routine.  {Specifically, $p^{\text{m}}(z_i)$ should be added at $z_j$ in Eq.~(\ref{eq:routine}).} With a constant $r_0$ and varying $\mathbf{B}_0$, we have obtained a series of stratifications with the SE ionization, as shown in Figure~\ref{fig:total_density}. We note that only the total densities of all fluids are shown for simplicity. Without magnetic pressure (gradient), the cHE stratification has a lower density than the averaged 2D solution. In comparison, adding magnetic pressure gradients significantly alters the solution, as stronger gradients (Figure~\ref{fig:ratio_comparison}) support higher density.  
More importantly, the results demonstrate that the current numerical integration routine can be used to calculate cHE stratifications under varying magnetic fields.

So far, the discussion only covers 1D scenarios with a constant gravitational acceleration. We might also consider stratification along a coronal loop, in which case the gravitational acceleration varies with the loop's curvature. This scenario is relevant for understanding such phenomena as wave heating \citep{Karampelas_2019}, MHD seismology \citep{Nakariakov2000}, or prominence formation \citep{Jercic_2025}. Without going into the details of realistic magnetic fields, we simply discuss a semicircular loop, for which a cosine function can be used in the integration routine straightforwardly, to revise the balance law between gravity and pressure gradient. We thus have 
\begin{equation} \label{eq:g_varying}
\frac{\text{d} P}{\text{d} z}=-\frac{p_{\text{T}}^{\text{g}}\overline{m}_{\text{T}}}{k_{\text{B}}T}g\cos\left(\frac{\pi z}{2L}\right), \quad z < L,
\end{equation}
 \noindent in which $L$ is the radius of the loop. This function is essentially the same as what was used by \citet{Karampelas_2019}. As $z$ is the height from the footpoint, the gravitational acceleration along the loop axis becomes zero at the loop apex ($z=L$).
 Plasma along a realistic loop is supported by plasma gas pressure and magnetic pressure/tension, and we can already see that magnetic pressure gradient effectively shapes the 1D stratifications in the chromosphere and above. 
 Overall, including Eq.~(\ref{eq:g_varying}) has a less significant effect on the stratifications, and corresponding results are thus not further discussed.

We note that it is possible to directly extract the required variables along open or closed field lines (or other regions of interest) from 2D and 3D solar atmospheric models, and then cHE stratifications can be constructed based on these variables.

\end{appendix} 

\end{document}